# Social Systems: Can We Do More Than Just Poke Friends?

Georgia Koutrika   Benjamin Bercovitz   Robert Ikeda   Filip Kaliszan
Henry Liou   Zahra Mohammadi Zadeh   Hector Garcia-Molina
Computer Science Department, Stanford University
353 Serra Mall, Stanford, CA 94305, USA
{koutrika, berco, filip.kaliszan, liouh, zahram}@stanford.edu
{rmikeda, hector}@cs.stanford.edu

## ABSTRACT

Social sites have become extremely popular among users but have they attracted equal attention from the research community? Are they good only for simple tasks, such as tagging and poking friends? Do they present any new or interesting research challenges? In this paper, we describe the insights we have obtained implementing CourseRank, a course evaluation and planning social system. We argue that more attention should be given to social sites like ours and that there are many challenges (though not the traditional DBMS ones) that should be addressed by our community.

## 1. INTRODUCTION

Social web sites, such as FaceBook, del.icio.us, Y! Answers, Flickr and MySpace, have gone from being a small niche of the Web to one of its most important components. In these sites, a community of users contribute resources, which can be photos, personal information, evaluations, votes, answers to questions or annotations. Social sites have become extremely popular among users but have they attracted equal attention from the research community? Or are they considered yet another type of web site or database application, where users do simple and uninteresting things, such as poking friends[1] and tagging photos? Do they present any new or interesting challenges to researchers?

Social sites *are different from the "traditional" open Web* in that each site is controlled by some entity that can set up "rules" of engagement. Also, these sites tend to foster communities of users that are authenticated in the system and regularly contribute resources. At the same time, social sites *are different from "traditional" database applications* in that the content is often unstructured and often multimedia, contributed by the users not by some "official" central source, and users may have fake or multiple ids. Also, in social sites, the "customers" have very diverse characteristics and goals, and the user experience is often paramount.

The strengths of the database community are on "back end" issues: achieving high transaction rates, optimizing complex SQL queries, or mining huge amounts of data. On the other hand, web research has focused on search and indexing technologies for unstructured data. While these are important issues in any system that handles large volumes of (structured or unstructured) information, they are not the ones that differentiate successful from less successful social sites. The special characteristics of social systems, which set them apart from classical systems, raise several important questions that remain unanswered:

- What are the most effective ways for users to interact: discussion forums, question/answer paradigms, tags?
- What can be shared among the users in a community and what is considered sensitive information?
- What information at these sites can be trusted? How can trust be built into or studied in a social site?
- What are the best ways for users to visualize and interact with information?
- How are resources used to interact with other users? What kind of interactions among users and resources can be defined?
- How do such systems evolve over time? How do resources, users, and their relationships change and how does this affect the whole user experience?

We believe that as time marches on, such "front end" issues will be more and more important, not just in social sites, but in any information management system. In this paper, we describe the insights we have obtained implementing CourseRank, an educational social site where Stanford students can explore course offerings and plan their academic program. Faculty members and university administrators can also participate, providing useful information for students. Although CourseRank was designed for Stanford, other universities have expressed interest, and we are exploring exporting CourseRank. In addition to offering a useful service to Stanford students, CourseRank provides an ideal platform for conducting hands-on research on social systems.

We start by describing the existing CourseRank system and we cover the "lessons learned" so far (Section 2). We believe that many of these lessons are not just applicable to a university system but to any social site. Specifically, as we will discuss, a corporate social site has many similarities with CourseRank. We discuss research challenges in CourseRank (and social sites in general) and our efforts (Section 3).

---

[1]Poking someone is a simple way to let someone know that you want to be friends [4].

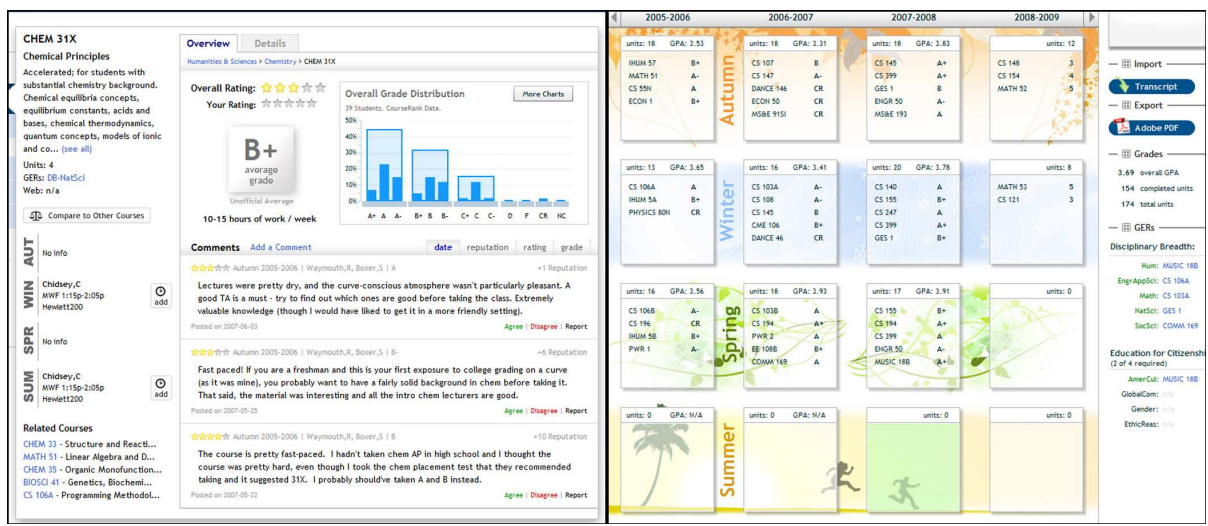

Figure 1: CourseRank Screen Shots: course description (left), course planner (right).

## 2. COURSERANK

Given the popularity of social sites, in our InfoLab at Stanford we decided (Summer 2007) to study information management in such sites. Although one can learn a lot by examining the handful of available traces and sample databases from commercial social sites, we quickly realized that without a site of our own it would be difficult to experiment with different algorithms and interfaces and do "out of the box" thinking. A lot of questions can only be answered by experimenting on a live system. In addition, if we wanted to attract users, we had to start with a niche area where we would not directly compete with the well established sites like FaceBook or Flickr. We set out to build CourseRank, a social site where Stanford students can review courses and plan their academic program by accessing official university information and statistics, such as bulletin course descriptions, grade distributions. Students can also provide information, such as comments on courses, ratings, questions and answers. To illustrate, the system provides (September 2008) access to $18,605$ courses, $134,000$ comments, and over $50,300$ ratings.

Using CourseRank, students can search for courses of interest, rank the accuracy of each others' comments and get personalized recommendations. They can shop for classes, and organize their classes into a quarterly schedule or devise a four year plan. Students can also check if the courses they have taken (or are planning to take) satisfy the requirements for their major. CourseRank also functions as a feedback tool for faculty and administrators, ensuring that information is as accurate as possible. Faculty can also modify or add comments to their own courses, and can see how their class compares to other classes. Figure 1 shows two CourseRank screen shots: on the left is part of a course descriptor page, and on the right is the 4-year course planner[2].

[2] At our site, (http://courserank.com), visitors can see a video with student testimonials and a demo (demo tab).

Initiated as a research platform, CourseRank was soon called "a long overdue success" (editorial in the Stanford student paper [6]). A little over a year after its launch, the system is already used by more than 9,000 Stanford students, out of a total of about 14,000 students. The vast majority of CourseRank users are undergraduates, and there are only about 6,500 undergrads at Stanford. Thus, CourseRank is already used by a very large fraction of Stanford undergrads.

### 2.1 Unique features

CourseRank has several important features that distinguish it from classical social sites but also from other public course evaluation sites (e.g., RateMyProfessors.com).

**Hybrid system**. CourseRank provides access to both official Stanford data (e.g., course descriptions, schedules and results of course evaluations conducted by the university) as in a typical database application, as well as to user-contributed information (e.g., course rankings, comments and questions) as in a typical social system.

**Rich data**. Courses, unlike books or videos, have to be taken in a certain order and in certain quarters. A course is offered by multiple instructors and may use multiple textbooks. Students enroll in courses and get grades. This richness of data introduces new challenges. For example, the recommendation system should take into consideration how useful a course is completing a major.

**New Tools**. In addition to providing tools similar to ones found at existing social sites (e.g., for searching for and evaluating courses), CourseRank offers powerful tools geared to our domain, for example, a tool for planning an academic program (*Planner*) that checks for schedule conflicts and computes grade point averages, a tool that checks if requirements for a major have been met (*Requirement Tracker*), and a tool for searching and browsing with help from a "tag cloud" (*CourseCloud*). CourseRank also offers a tool for "flexible recommendations" (*FlexRecs*) for the site administrator. This tool lets the administrator quickly define recommendation strategies that can be then selected (and person-alized) by a student who needs recommendations. Figure 2 sketches the several components that comprise CourseRank. In Section 3, we describe the CourseCloud and FlexRecs.

|          | DB                              | Web                               | Social Sites                  | CourseRank                      |
|----------|---------------------------------|-----------------------------------|-------------------------------|---------------------------------|
| **data** | centrally controlled            | uncontrolled, highly distributed  | centrally stored              | centrally stored                |
|          | transactional, "official"       | many providers                    | user contributed              | user contributed + official     |
|          | structured                      | unstructured + deep web           | mostly unstructured           | both types                      |
|          | very large                      | humongous                         | extra large                   | large                           |
| **access** | 1 provider − many consumers   | many providers − mass consumers   | users-to-users                | closed community                |
| **users** | authorized                     | anyone                            | authorized                    | authorized                      |
|          | real ids                        | anonymous                         | fake and multiple ids         | real ids                        |
|          | very focused interests          | diverse interests (hard to know)  | shared but diverse interests  | community-shaped interests      |
| **apps** | financial                       | keyword search                    | bookmarking                   | university site                 |
|          | telecommunications              | browsing                          | networking                    | corporate site                  |
| **research** | long-time established       | index and search                  | little research               | lots of challenges              |
|          | ACID database                   | little db technology              | home-made solutions           |                                 |

Table 1: Comparing CourseRank to Social Sites to Classical Systems

**Site Control**. Unlikely the "open" Web, all data is centrally stored and we have control over the site.

**Closed Community**. CourseRank is only available to the Stanford community.

**Constituents**. Many social sites have a single type of user, although they can sometimes be divided into "power users" (e.g., who store many photos and perhaps pay a fee) and regular users, and sometimes they can be divided by geography and interest (e.g., in FaceBook users join one or more networks). In CourseRank, there are three very distinct types of users: (a) *Students* (undergraduate and graduate); (b) *Faculty* who may want to check comments on their courses and compare against other courses; and (c) *Staff*, who can enter requirements for academic programs, and who may want to advise a student on her course plan.

**Restricted Access**. CourseRank has access to official "user names" on the Stanford network and can therefore validate that a user is a student or a professor or staff.

Table 1 summarizes the features and differences of DB applications, the open Web, social sites, and CourseRank. CourseRank's unique features provide both opportunities and challenges, as we comment when we discuss the lessons we have learned and the initial research we have conducted.

## 2.2 Lessons Learnt so Far

What makes a social site successful? Here is what we have learned from building and running CourseRank over the past year.

**Meaningful Incentives**. In a social site, there need to be incentives for users to visit and to share their resources. The incentives are especially critical in the early stages, where there are few resources shared by others. Many sites use mechanisms to incentivize their users. For instance, Yahoo! Answers [7] uses a scoring scheme: providing a best answer is rewarded by 10 points, logging into the site yields 1 point a day, voting on an answer that becomes the best answer increases the voter's score by 1 point, and so forth. However, such incentives do not necessarily make users contribute sensibly. Users often try to boost their reputation by exploiting these schemes.

Providing meaningful incentives to our users was very important. Students provide personal information (e.g., their class, major), the courses they have taken and their grades, because they can use tools, such as the course planner and the calendar, to help them structure their courses over multiple years. For instance, the planner has been an extremely useful feature, so users have a reason to visit beyond just looking for courses to take. It is also a sticky feature. Once a student has taken the time to enter his courses and grades, he keeps returning. The planner motivates the student to enter accurate data: since it shows to its owner grade averages per quarter, and missing requirements for graduation, there is little reason to lie about courses taken.

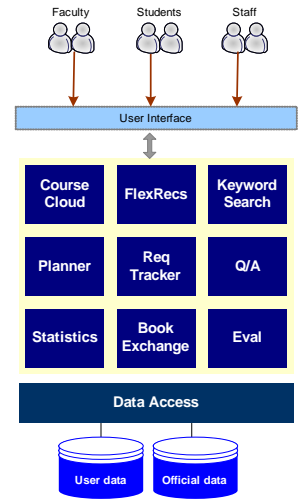

Figure 2: CourseRank System.

Interestingly, not all features have been popular to date. Our Question and Answer forum has little traffic because there are no incentives to visit: If there are few questions or answers, why would people ask questions or go looking for answers there? To address this shortcoming, we plan to seed the forum with "frequently asked questions" developed in conjunction with department managers, e.g., "who do I see to have my program approved?" or "what is a good introductory class in department X for non-majors?" Questions will be automatically routed to people who are likely to be able to answer them. With a useful body of questions and answers, we hope students will start using the forum.

**Interaction for Constituents**. Because of our authentication system, we know the type of a user. We have learned that it is important to offer specialized features for each of our constituencies, to motivate them to use our site and provide resources that can help the other constituencies. For example, we provide a dedicated interface for department managers that allows them to define the requirements for their programs. This interface has the potential to reduce their workload, so they are happy to work with us. At the same time, having the requirements entered in the system enables students to check which requirements they meet based on the courses they have taken so far. We also offer special features for faculty members to enter information on their courses, such as updates to the official course description and pointers to other useful materials that may help students decide if the course is for them. For a successful system, we feel it is critical to have incentives for each constituency, and to meet their differing requirements.

**The Power of a Closed Community**. CourseRank is a social site for a small, closed community, where users have known identities. We believe this usage differs greatly from what is seen in general-purpose social sites that are open to anyone and, hence, may attract spammers and malicious users. In CourseRank, users are more willing to contribute more thoughtfully. For example, not only we have a substantial body of comments but in the data we have for the first 12 months of operation, we already see much higher quality comments than what one typically finds in public course evaluation sites or in social sites. Users in CourseRank put more effort to contributing to the system, and as an implicit consequence of this effort, they trust the system.

**It's the Data, Stupid**. While some of the data in CourseRank is entered by users (course evaluations, courses taken, self reported grades), one key to its success was the availability of useful external data, such as course descriptions and schedules, associated textbooks, official grades distributions, and so on. Having official data in combination with user input added value to the system.

Many sites of course depend on external data: goods for sale, the news, weather reports, etc. Getting the physical bytes is the easy part, getting permissions and understanding what can be done with the data is the hard part. Of course, extracting, transforming and loading data, ETL, is a hard problem, but the data warehousing community has developed useful techniques for that side of the problem. Getting the rights to use the data is harder still, since it involves continuous negotiations.

In some cases, the issues are economic. For example, our own Stanford Bookstore did not want to release the list of textbooks associated with each class, even though the information came from our own professors, and our goal is only to *help* Stanford students by giving them more options for buying textbooks. Instead we had to implement a system for volunteers to report textbooks to CourseRank, which is working very well. Thus, even though the information is public, the people holding the information have a financial stake in it and are not willing to share it.

In other cases, the issues revolve around privacy. For example, the distribution of grades (not individual grades) in each class has for years been available to Deans and department heads. However, each professor only got to see his or her own distribution (which of course he already knew!), and even the distribution of grades within a department was a closely guarded secret for some departments. Of course, students have always known what the "easy courses" are, and now with CourseRank they were able to see the distribution of the self-reported grades. We have argued that it is better to disclose the true distribution, than our approximation, but so far only the School of Engineering has bought our argument. Thus, we now display the official distribution only for engineering courses. Of course, we do not show distributions for classes with very few students, since that may disclose information about individual students. (Incidentally, the official Engineering grade distributions seem to be very close to the corresponding self-reported ones, validating our claim that students are entering valid data.)

In summary, a social site needs interesting high-quality data. Access to it needs to be carefully negotiated with the owners, since people are very attached to their data (even if the data reflects public knowledge).

**Privacy can be "shared"**. While our data providers were very concerned about their data, and one frequently hears about privacy concerns in the news, our users are actually unconcerned about privacy in many cases. We think there are two reasons for this. CourseRank can only be accessed by Stanford students. Thus, students know the data they make available will only be seen by other students like them. The second reason is that the students are young. As anyone who has visited the pages of young people on FaceBook or MySpace, young people have different standards as to what is sensitive. These popular social sites seem to be changing the culture, and their members are much more willing to share information and to communicate in public (e.g., using "walls" in FaceBook).

We offer one anecdote to illustrate the more open nature of CourseRank. In our tool, students add courses they are planning to take (in addition to the ones they have already taken). The oldest member of the CourseRank team thought this information was clearly sensitive and should not be disclosed. However, the feedback we received was that students wanted this information to be shared: If Sally knows that Bob is taking CS106, and Sally likes Bob, then Sally can enroll in the class with Bob! Thus, we allowed students to see who is planning to take a class (one can opt out of sharing), and this feature has been very successful. Apparently the vast majority of students do not view their plans as sensitive.

**Closed Loop Feedback**. CourseRank was mainly implemented by three Stanford undergraduates, and what is interesting is that the students and faculty like CourseRank much better than most of the official university sites. For example, the university recently released a site for our Career Placement Center, and the student paper asked in an editorial why this system was so poor, having been built by an outside contractor for a substantial sum of money, while CourseRank was so good, having been implemented by students. Similarly, some of the official course evaluation data released by the university has also been placed on the Stanford administrative site (Axess), but the data is hard to find and understand, while it is much easier to find and much better graphed in CourseRank.

How can CourseRank compete so well with professionally developed systems? One reason is that the developers (students) are intimately familiar with the application (evaluating courses, planning for their degree), so they know what features will be useful and how other students want to use these features. Furthermore, there is a tight feedback loop with "customers." Our developers lived with actual and po-tential users, so they constantly got feedback. Many of our features were suggested or significantly improved by user feedback. In summary, when the selling point of a system is the user experience (as opposed to how fast the back end is), it is critical to know the application well and to be constantly receiving feedback from customers.

**Beyond CourseRank: The Corporate Social Site.** We believe that many of these lessons are not just applicable to a university system but to any social site. In particular, we envision a corporate social site where employees and customers can interact and share experiences and resources. A corporate site shares many features with CourseRank: the need to service a varied constituency (employees, managers, customers, etc), restricted access, having the control of the site, and so forth. Several companies are tracking our progress on CourseRank, with an eye to building a corporate site of their own.

## 3. INTERACTION WITH RICH DATA

As we have argued, CourseRank provides an excellent testbed for studying social systems and for identifying needed features. Interestingly, we have access to much richer data than a typical social site, such as course information, user profiles (major, class, grades), course interrelationships (e.g., courses needed for major, pre-requisites), and so on. Interacting with rich data calls for powerful models and methods.

One of the most frequent requests we have got is for more powerful search and discover mechanisms. Stanford offers a wealth of courses (18,605), but it is hard to navigate through them all. Search engines are good at finding things once we know the important keywords. Traditional browsing schemes (say of a course catalog) are not good for making unexpected connections (serendipity). For example, a student browsing through the listings of the classics department looking for "something" related to Greece may not find a course on the history of science that covers some of the famous Greek scientists. Of course, if she knew the keywords "Greek, science", she could find it, but she may not initially have made the connection.

On the other hand, traditional recommendation engines can select popular items, e.g., popular courses or courses that "people like me" have taken, but they offer little control. For instance, a student may want to base her recommendations on people with similar grades, as opposed to with similar tastes. Or maybe a student is not looking for a course, but is looking for a major that suits the courses she has taken, or trying to figure out what is the best quarter to take a calculus course this year. How can different types of information be combined to provide more expressive and targeted recommendations? For example, could we take into account the student's personal interests and also grade history to recommend appropriate courses? Interrelationships between different types of data should be taken into account. For instance, if a course $A$ has as a prerequisite a course $B$, then $A$ should not be recommended independently.

In this section, we briefly describe two new features we added to CourseRank to address these challenges: Data Clouds and Flexible Recommendations.

### 3.1 Data Clouds

A data cloud is a tag cloud, where the "tags" are the most representative or significant words found in the results of a

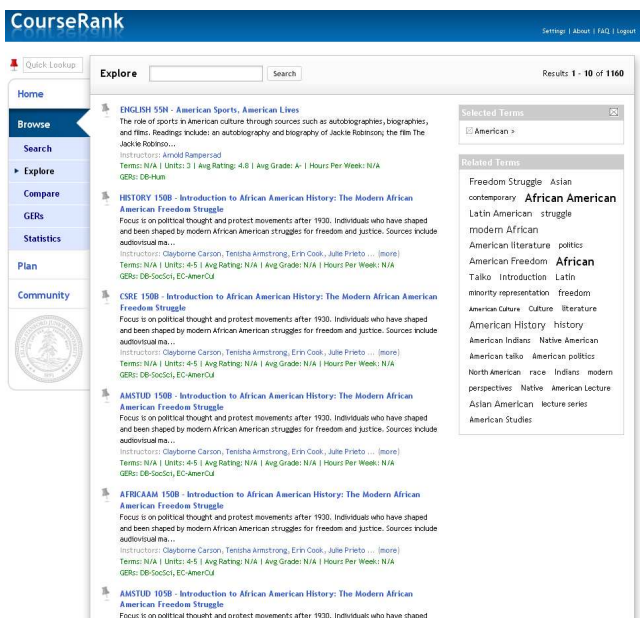

Figure 3: Searching for "American".

keyword search over the database. Data clouds summarize search results and help users refine their searches. In CourseRank, a data cloud is used to summarize the results of a keyword search for courses, and is called course cloud. We could easily expand searching with clouds to other entities, such as books and instructors.

For instance, a student interested in taking a class on American culture and history can type "American" and get a list of matching courses along with a cloud summarizing course information in this list, as shown in Figure 3. The keyword "American" is searched in different fields and relations in CourseRank's database, not just in the main course relation. For example, if there are comments that mention "American", the respective courses will appear (in some position) in the results. There are 1160 courses returned for this search. The cloud provides related concepts that are found in the matching courses, such as "Latin American", "Indians", and "politics". These words may be found in different parts of the database related to the current search. For example, the term "Latin American" may also exist in user comments that refer to American-related courses.

Terms in the data cloud (as in traditional tag clouds) are hyperlinks. The searcher can click on a term from the data cloud to refine search results. For example, she can click on "African American" and narrow down the results to 123 matches. The cloud is updated accordingly to reflect the new, refined, results. Figure 4 shows the updated list of courses and the respective cloud. Different students may choose different terms from a data cloud refining their searches in diverse ways.

Hence, we couple the flexibility of keyword searches over structured data with the summarization and visualization capabilities of tag clouds to help users search a database. Data clouds can illustrate "interesting" terms in the database and can lead to serendipitous discoveries. At the same time, they raise important questions:

- How do we effectively define and search over search enti-

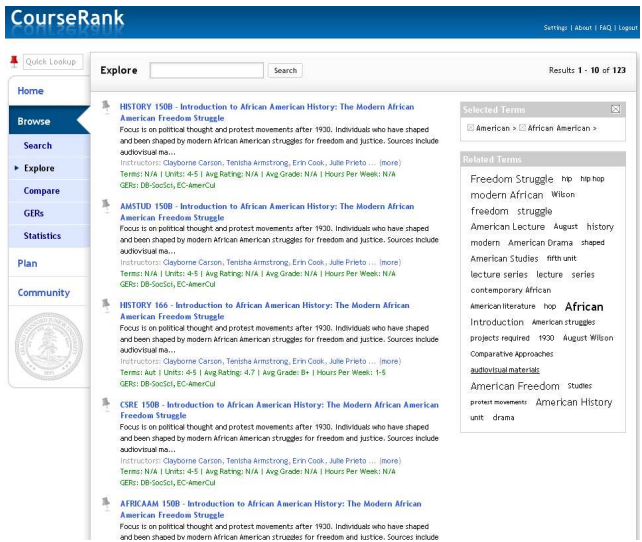

**Figure 4: Searching for "African American".**

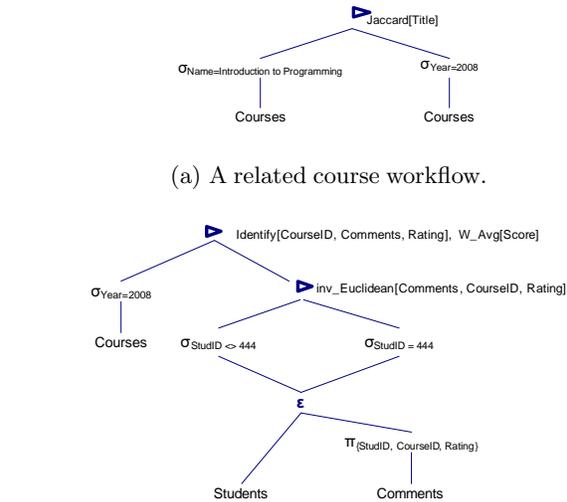

(a) A related course workflow.

(b) A collaborative filtering workflow.

**Figure 5: Two sample workflows.**

ties that span multiple relations rather than over tuples as in traditional database querying? For instance, we may want to define a course entity to include not just its title and description, but all the comments made by students about the course, and the names of all the students that are enrolled in the course.

- How do we rank search entities depending on the position of a query term? If we search for "`Java`" courses, should a course that mentions "`Java`" in its title have the same score as a course that mentions "`Java`" in the comments made by students about the course?

- The data cloud contains the most significant or representative terms within the currently found set of entities. The terms are aggregated over all parts that make a course entity and may be stored in multiple relations. How do we find and rank terms in the results of a search and how can we dynamically and efficiently compute their data cloud?

## 3.2 Flexible Recommendations

Most recommendation systems have several limitations. They model the world as having two types of entities, users and items (e.g., movies), represented as sets of ratings or features. Providing recommendations using richer data representations is not straightforward. For example, a user may want recommendations for *courses* from users *with similar grades and similar ratings*. Furthermore, the recommendation algorithm is typically embedded in the system code. From the designer viewpoint, it is hard to modify the algorithm, or to experiment with different approaches. The recommendations provided are typically fixed and end users are given few choices. For example, a user may be unable to request that her recommendations be based on what people *in her major* are taking rather than based on all students.

We are implementing a novel engine, *FlexRecs*, that allows *flexible recommendations* to be easily defined, customized, and processed. A given recommendation approach can be expressed *declaratively* as a high-level workflow over structured data. At the heart of FlexRecs lies a special *recommend operator*, which takes as input a set of tuples and ranks them by comparing them to another set of tuples. The operator may call upon functions in a library that implement common tasks for recommendations, such as computing the Jaccard or Pearson similarity of two sets of objects. The operator may be combined with other recommend operators and traditional relational operators, such as select and join operators.

For instance, suppose that our information on courses, students and evaluations is stored in the following relations:

Courses(<u>CourseID</u>, DepID, Title, Description, Units, Url)
Students(<u>SuID</u>, Name, Class, GPA)
Comments(<u>SuID</u>, <u>CourseID</u>, <u>Year</u>, <u>Term</u>, Text, Rating, Date)

Assume we are interested in finding courses for 2008 that are similar to the course with title "Introduction to Programming". Figure 5 shows two workflows one can define with FlexRecs. In the workflow shown in Figure 5(a), we simply find courses with titles similar to the indicated course. The workflow shown in Figure 5(b) uses two recommend operators. The lower recommend operator (triangle) first computes the students that are similar to a target student with id 444. Similarity between students is computed by taking the inverse Euclidean distance of their ratings. The second recommend operator (upper triangle) uses the similar students to rank courses. Specifically, a course's score is the average of the ratings given by the similar students. The workflow additionally contains an *extend* operator ($\varepsilon$), which allows the recommend operator to view the set of ratings for each student as another attribute of the student irrespective of the database schema.

Decoupling the definition of a recommendation process from its execution allows defining easily new recommendation types and executing them by the same engine. The engine executes a workflow by "compiling" it into a sequence of SQL calls, which are executed by a conventional DBMS. When possible, library functions are compiled into the SQL statements themselves; in other cases we can rely on external functions that are called by the SQL statements.

FlexRecs let us experiment with different recommendation strategies (workflows), and offer users options for personalizing recommendations. In CourseRank, we are implementing an interface where one can ask for recommended courses, or recommended majors (for students that have not declared a major), or recommended quarters in which to take a given course and choose different options on how recommendations will be generated (e.g., based on what "similar" students have done or the grades they have taken). There are many challenges ahead:

- Handling the full suite of FlexRecs operators is more challenging than what our simple examples illustrate. How can we optimize the execution of workflows? How can we implement the operators extending appropriately the database engine?

- What is an appropriate interface for allowing users to control recommendations? Different levels of user involvement will make sense for different applications. For example, students in CourseRank may require more flexibility than customers of an online movie renting service.

## 4. RELATED WORK

A growing number of social systems can be found on the web enabling people to share different kinds of resources, such as: photos (e.g., Flickr [3]), URLs (e.g., Del.icio.us [2]), blogs (e.g., Technorati [5]), research papers (e.g., CiteULike[1]), and so forth. CourseRank is one of the first social system, whose purpose goes beyond resource sharing and networking. Instead, it enables course planning through a number of social activities and interactions, such as sharing, rating, and so forth.

Research on social sites has mainly focused on understanding the usage and evolution of these systems [8, 10, 11, 13, 15, 19, 20]. As far as we know, there has been little research in our community (and other communities for that matter) on social information management aspects, such as interacting with rich data. There is a small number of efforts on harvesting social knowledge in social networking systems for different purposes, including *resource recommendations* [16, 17, 20], *expert and community identification* [9, 12, 14, 21] and *ontology induction* [18].

## 5. CONCLUSIONS

In this paper, we have argued that more research attention should be given to social sites. In particular, we envision social sites that can be used to support a well-defined community, in a university or in a corporation. We have presented CourseRank as an example of such a *focused* social site. A focused social site will typically have a closed community, which will make users more willing to contribute. It will also have rich data that requires improved social interaction tools. We provided two examples of such tools: Data Clouds for information discovery through tag clouds, and FlexRecs for declarative definition of recommendation strategies.

*In conclusion*: social sites are not just for sharing links and networking. They can provide valuable services based on user contributed information, and they present interesting information management and interaction challenges.